\documentclass[fleqn,10pt]{wlscirep}
\usepackage[utf8]{inputenc}
\usepackage[T1]{fontenc}
\usepackage{bm}

\usepackage{amsbsy}

\usepackage{color}

\title{Collective motion of active particles exhibiting non-reciprocal orientational interactions}

\author[1,*]{Milo\v s Kne\v zevi\'c}
\author[1]{Till Welker}
\author[1]{Holger Stark}
\affil[1]{Institut f\"ur Theoretische Physik, Technische Universit\"at Berlin, Hardenbergstra\ss e 36, D-10623 Berlin, Germany}

\affil[*]{knezevic@campus.tu-berlin.de}

\begin{abstract}
We present a Brownian dynamics study of a 2d bath of active particles interacting among each other through usual steric interactions and, additionally, via non-reciprocal avoidant orientational interactions.
We motivate them by the fact that the two flagella of the alga \emph{Chlamydomonas} interact sterically with nearby surfaces such that a torque acts on the alga.
As expected, in most cases such interactions disrupt the motility-induced particle clustering in active baths. Surprisingly, however, we find that the active particles can self-organize into collectively moving flocks if the range of non-reciprocal interactions is close to that of steric interactions. We observe that the flocking motion can manifest itself through a variety of structural forms, spanning from single dense bands to multiple moderately-dense stripes, which are highly dynamic.
The flocking order parameter is found to be only weakly dependent on the underlying flock structure.
Together with the variance of the local-density distribution, one can clearly group the flocking motion into the two separate band and dynamic-stripes states.

\end{abstract}
\begin{document}

\flushbottom
\maketitle
\thispagestyle{empty}

\section*{Introduction}

\begin{figure}
	\centering
\includegraphics[width=15.5cm]{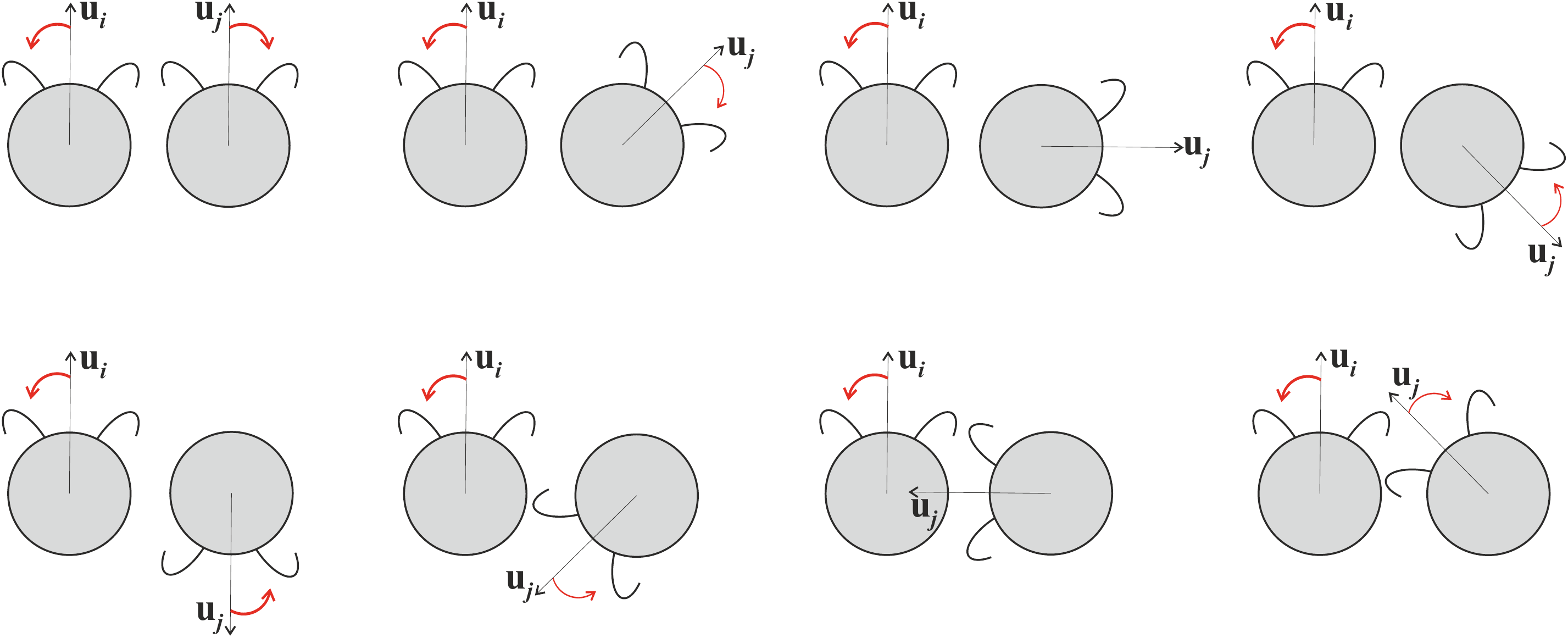}
	\caption{A schematic of the encounter between two \textit{Chlamydomonas}. Their current unit orientation vectors $\mathbf{u}_i$ are depicted in light black, while their directions of rotation upon interactions modeled by non-reciprocal torques from Eq. (\ref{eq:nonrec}) are displayed in red. As an illustration eight different examples are shown. The torques are non-reciprocal in the general case (hinted by different thicknesses of the red arrows).
	}
	\label{fig:1}
\end{figure}  

The interdisciplinary field of active matter\cite{ramaswamy:10,vicsek:12,elgeti:15,zoettl:16,bechinger:16}, situated at the crossroad of physics, chemistry, biology and engineering, examines individual and collective behavior of units capable of autonomous motion. The units can be diverse, including, for instance, bacteria that use body appendages to swim in viscous media, active colloids that exhibit propulsive motion by generating flow around their surfaces by means of self-phoresis, bird flocks and sheep herds. Many interesting properties of these systems can be captured already with simple models which do not account explicitly for the medium surrounding active units. Instead, on the coarse-grained level the medium serves as a source of fluctuations and drag forces acting on the units. Typically, active units are modeled as particles moving with a roughly constant speed along an orientation vector attached to them\cite{romanczuk:12}, which can change its direction due to rotational diffusion and inter-particle interactions. On one side of the spectrum, there are models with pairwise interactions. Possibly the simplest example within this class of models involves isotropic active particles interacting with each other via steric interactions, which can lead to the phenomenon of motility-induced phase separation\cite{tailleur:08,filly:12,redner:13,cates:15} (MIPS) between dense particle clusters and dilute regions of freely moving particles. In a nutshell, MIPS takes place when active particles are sufficiently fast and numerous\cite{buttinoni:13,speck:16}, increasing chances that free particles join existing clusters compared to chances that particles within clusters escape them as a result of rotational diffusion. On the other side, there are models with social interactions\cite{lavergne:19,baeuerle:20}. The most notable example within this class is the Vicsek model\cite{vicsek:95}, incorporating alignment with average particle orientation of all neighbors situated within some fixed radial distance, which can result in coherent flocking motion\cite{toner:95,czirok:97,toner:98,gregoire:04}: particles spontaneously form traveling bands propagating along a common direction. Models fusing both type of interactions\cite{martin-gomez:18} are also found in literature.
Furthermore, different types of orientational interactions have been introduced \cite{Zhang:21,Linden:19,Pu:17,Liao:20,Barre:15,Sansa:18,Bhatta:19, Mallory:19}. In this article we introduce a specific form of orientational interactions, which are non-reciprocal and avoidant. We demonstrate how they influence the collective dynamics of our model active particles. The orientational interactions are motivated by the fact that the two flagella of the alga \emph{Chlamydomonas} interact sterically with nearby surfaces such that the alga experiences a torque, which rotates it away from the surface.

If pairwise interactions stem from a potential, they are reciprocal due to Newton's action-reaction principle. However, the reciprocity can be broken if the pairwise interaction originates in some non-equilibrium mechanism\cite{ivlev:15}. Numerous examples exist in literature, including carpets of microfluidic rotors\cite{uchida:10}, droplets exhibiting predator-prey interactions driven by non-reciprocal oil exchange\cite{meredith:20}, neural networks\cite{sompolinsky:86,montbrio:18} and many others\cite{fruchart:21}. An eminent example of non-reciprocal interactions from the realm of active matter is found in binary mixtures of active colloids with phoretic interactions
\cite{soto:14,pohl:14,saha:19,stuermer:19}. In essence, colloids with two distinct hemispheres, immersed in a solution containing an appropriate chemical solute, can produce and maintain a local chemical concentration gradient by means of, for instance, self-diffusiophoresis or self-thermophoresis. The chemical gradient implies a hydrodynamic flow of the surrounding solvent, which in turn propels the colloid forward so that the entire system remains force free. Two populations of different active colloids will exhibit a different response to self-generated chemical gradients, leading in general to non-reciprocal interactions, the strength of which can be controlled by varying colloidal mobilities and activities. Interestingly, it has been reported that interplay of hydrodynamics and boundaries can also lead to non-reciprocal interactions\cite{uchida:10}.

In this article we study the collective motion of a 2d low Reynolds number suspension of self-propelled particles displaying non-reciprocal pairwise interactions. The particles perform active Brownian motion with a constant swim speed $v$ and translational and rotational diffusion constants $D$ and $D_\mathrm{R}$. Two types of pairwise interactions are present in the system as we describe in more detail in the Methods section. Firstly, the particles interact with each other via repulsive steric forces with a typical 
interaction range $\sigma$, which can be interpreted as an effective particle diameter. In the absence of other kind of interactions, the active suspension would be described by two dimensionless parameters: the area fraction $\Phi$ of active particles and the P\'eclet number $\text{Pe} = v\sigma/D$. The $\text{Pe}$ number is the ratio of characteristic times $\sigma^2/D$ and $\sigma/v$ that the active particle needs to respectively diffuse and swim its own size $\sigma$. In this study we fix $\Phi = 0.24$ and $\text{Pe} = 80$ such that the system is in the regime where MIPS occurs.   

In addition, the particles exhibit pairwise orientational interactions of short range $R$, which manifest themselves as torques acting on the particles.
Motivated by the avoidant torque, the alga \emph{Chlamydomonas} experiences close to a surface as explained below, we have formulated the orientational interaction of  Eq. (\ref{eq:nonrec}) in the Methods section.
If the distance $\mathbf{r}_{ij}$ between the particles $i$ and $j$ is smaller than a cutoff distance $R$, they are subjected to torques of the amplitude $\mathcal{T}_0$. In the general case, the orientational interaction is non-reciprocal, meaning that the torques $\pmb{\mathcal{T}}_{ij}$ and $\pmb{\mathcal{T}}_{ji}$ 
acting on the particles are not equal and opposite, $\pmb{\mathcal{T}}_{ij} \neq -\pmb{\mathcal{T}}_{ji}$. This stems from the form of Eq. (\ref{eq:nonrec}) where the torque exerted on particle $i$ depends on its orientation $\mathbf{u}_i$ and the normalized separation vector $\mathbf{r}_{ij}/|\mathbf{r}_{ij}|$, but not on the orientation of particle $j$. The interactions are predominantly avoidant in nature. For example, interaction of such form can be motivated by looking at the dynamics of alga \textit{Chlamydomonas reinhardtii}\cite{harris:09}. \textit{Chlamydomonas} propagates itself through the fluid by breaststroke 
beating of a pair of frontal flagella\cite{polin:09} such that its far-field flow topology resembles that of a puller\cite{drescher:10}: the flow is inward along the main body axis and outward in the perpendicular direction. Recent experiments have demonstrated that \textit{Chlamydomonas} scatter off solid surfaces primarily due to contact interactions\cite{kantsler:13} between their flagella and the surface, while later investigations reveal that hydrodynamic interactions are needed for a complete quantitative description of how algae interact with cylindrical obstacles \cite{contino:15}.
The contact interactions, on which we concentrate here, generate torques\cite{ostapenko:18} that avert the algal body from a collision with the surface. To capture this type of interaction, we propose the model described by Eq. (\ref{eq:nonrec}), in which case $\pmb{\mathcal{T}}_{ij} \equiv \pmb{\mathcal{T}}_{i}$ would stand for the torque exerted on the particle and $\mathbf{r}_{ij} \equiv \mathbf{r}_i$ for the distance vector of the particle $i$ from the surface. It is now reasonable to suppose that an analog contact interaction through flagella exists among two algae situated close to each other. Since there are no useful experimental insights concerning such interactions between algae, we propose the torque of Eq.\ (\ref{eq:nonrec}) as a first approach for describing these interactions, which certainly needs to be refined in future work for a quantiative description. Thus, as the flagellum of one \textit{Chlamydomonas} touches the cell body of the other, and vice versa, the torques exerted on the algae will be non-reciprocal in the general case. Some typical examples of the encounter of two algae modeled by Eq. (\ref{eq:nonrec}) are illustrated in Fig. \ref{fig:1}, where the orientation of alga $i$ is fixed and eight orientations of alga $j$ are chosen. The directions of the torques acting on both algae and their strengths are indicated by curved red arrows.
In particular, the avoidant torque on particle $j$ is maximal if its orientation vector $\mathbf{u}_j$ is perpendicular to the distance vector to particle $i$ ($\mathbf{u}_j \perp \mathbf{r}_{ij}$) and, by symmetry, zero for $\mathbf{u}_j \| \mathbf{r}_{ij}$. The torque in Eq. (\ref{eq:nonrec}) can be viewed as a simple and generic form of an avoidant non-reciprocal orientational interaction and, therefore, in the following we explore how it influences the collective motion of active Brownian particles.

\begin{figure}
	\centering
	\includegraphics[width=14cm]{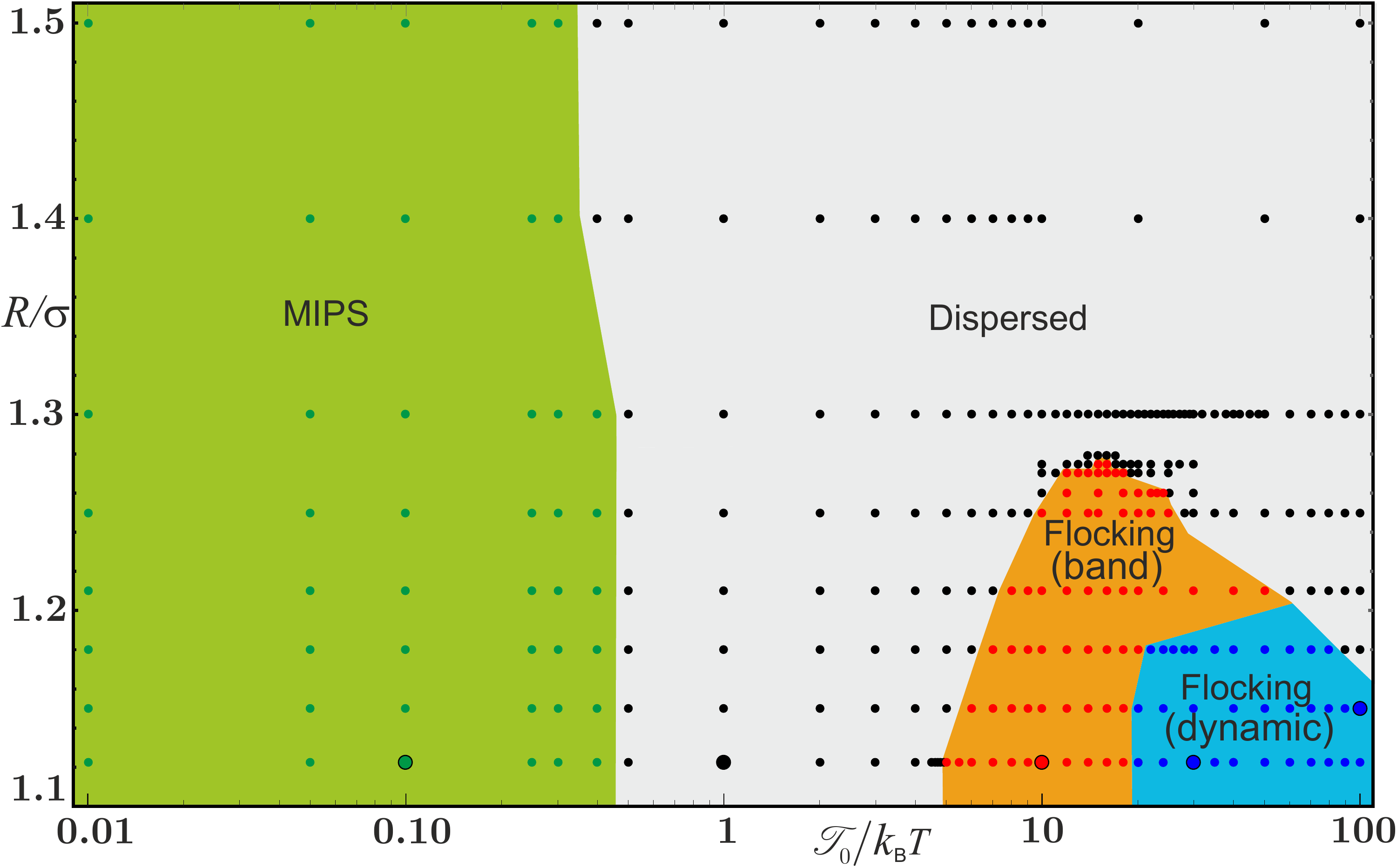}	
	\caption{State diagram of the system as a function of the non-reciprocal interaction range $R$ and interaction amplitude $\mathcal{T}_0$. The $\mathcal{T}_0$ axis is presented in the logarithmic scale and includes data over the range of five orders of magnitude. Apart from the dispersed state, three ordered states are observed, which we term MIPS, flocking (band) and flocking (dynamic).
	State boundaries have been sketched by using simulation data, represented by circles of different colors. The system is characterized by $\Phi = 0.24$ and $\text{Pe} = 80$. The states of the snapshots presented in Fig.\ \ref{fig:3} are indicated by enlarged circles.	
	}
	\label{fig:2}
\end{figure} 

We describe active particle motion by a set of coupled overdamped stochastic equations, which are presented in the Methods section. Starting from an active suspension exhibiting MIPS as a reference state, we introduce non-reciprocal orientational interactions (\ref{eq:nonrec}) and using the method of Brownian dynamics conduct a detailed parameter study of collective motion of the resulting active bath as a function of torque amplitude $\mathcal{T}_0$ and interaction range $R$. As the orientational interactions are predominantly avoidant, one expects disruption of MIPS for strong enough torques $\mathcal{T}_0$, because active particles at the edge of a cluster turn away from neighboring particles and leave the
cluster.
Even though the disordered state is indeed the most common outcome for strong torques, as the state diagram in Fig.\ \ref{fig:2} shows, surprisingly, we find that there exists a narrow region in the state diagram, characterized by short interaction range $R$ and moderate torques $\mathcal{T}_0$, where a transition to ordered collective motion occurs. In the ordered state, the particles move in flocks which can take a variety of structures spanning from a compact dense band to a single very dynamic porous band and multiple bands. This is the main result of our study.

The article is organized as follows. In the Results section we analyze different states appearing in our system and construct the corresponding state diagram in the parameter space of $\mathcal{T}_0$ and $R$. We elaborate the implications of our findings, suggest possible future experiments, and offer our conclusions in the Discussion section. Lastly, the equations of motion of active particles and details of Brownian dynamics simulations are presented in the Methods section.

\begin{figure}
	\centering
	\includegraphics[width=16cm]{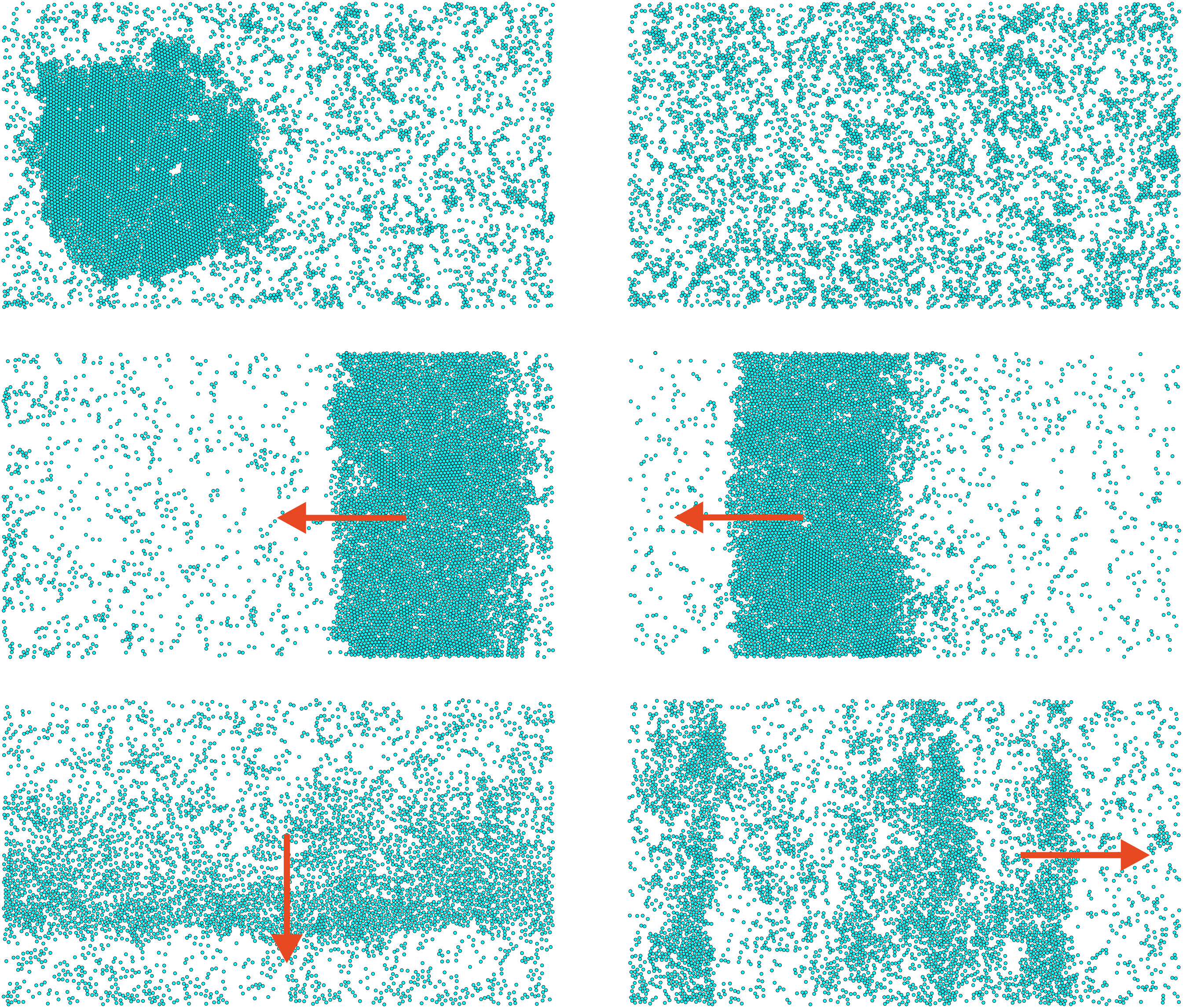}
	\caption{Simulation snapshots of typical states appearing in the system for different interaction strength $\mathcal{T}_0$ and range $R/\sigma$ as indicated by enlarged circles in the state diagram of Fig.\ \ref{fig:2}. Top left panel: MIPS ($\mathcal{T}_0/k_\mathrm{B}T = 0.1$ and $R/\sigma = 2^{1/6}$) and top right panel: dispersed state ($\mathcal{T}_0/k_\mathrm{B}T = 1$ and $R/\sigma = 2^{1/6}$). 
	Middle panel: Flocking (band). Two snapshots of a coherently moving band of active particles ($\mathcal{T}_0/k_\mathrm{B}T = 10$ and $R/\sigma = 2^{1/6}$) are shown, with a time of $3D_\mathrm{R}^{-1}$ in between them. Bottom panel: Flocking (dynamic).
	Left: Motion of a moderately dense and dynamically varying stripe of particles ($\mathcal{T}_0/k_\mathrm{B}T = 100$ and $R/\sigma = 1.15$). Right: Motion of several dynamically varying stripes ($\mathcal{T}_0/k_\mathrm{B}T = 30$ and $R/\sigma = 2^{1/6}$). 
	Red arrows denote the direction of coherent motion. In all cases $\Phi = 0.24$ and $\text{Pe} = 80$.
	}
	\label{fig:3}
\end{figure} 

\section*{Results}

The results of our Brownian dynamics simulations are summarized in a state diagram shown in Fig. \ref{fig:2}. We observe four distinct states.
First, for relatively weak torques, $\mathcal{T}_0/k_\mathrm{B}T \lessapprox 0.5$, non-reciprocal interactions play only a minor role and the active bath exhibits MIPS irrespective of their range $R$. This regime is illustrated in the top left panel of Fig. \ref{fig:3} and in Movie 1. Second, for sufficiently strong torques, $\mathcal{T}_0/k_\mathrm{B}T \gtrapprox 0.5$, and large interaction range $R$ the non-reciprocal interactions prevent motility-induced clustering. A typical snapshot of the resulting dispersed state is shown in the top right panel of Fig. \ref{fig:3}. Smaller unstable clusters are visible (see Movie 2).

The most interesting behavior occurs when the range of non-reciprocal interactions is close to the range of steric interactions, in our case $R/\sigma \lessapprox 1.3$. For moderately large torques, $\mathcal{T}_0/k_\mathrm{B}T \gtrapprox 5$, one observes a transition to coherent flocking motion of active particles. Depending on the exact parameter choice from this domain, the system can display an assortment of different flocking structures, which we roughly classify into two groups, which we will justify quantitatively below.
In the region displayed in orange in Fig. \ref{fig:2} a dense band of particles develops, cruising collectively in a common direction against a dilute background. This state is depicted in the middle panels of Fig. \ref{fig:3}. There, a coherently moving band is shown at two time instances, $3D_\mathrm{R}^{-1}$ apart from each other; see also Movie 3. In addition to ordered bands, other structurally diverse states displaying coherent motion are possible in the system. Some examples are shown in the bottom panels of Fig. \ref{fig:3}, which include a moderately dense stripe and multiple stripes. The dynamics of multiple stripes is shown in Movie 4. Other irregular structures are also possible.
They share the property that in contrast to the moving band, which always keeps its integrity, the stripes are very dynamic. As we demonstrate now, they can be grouped into one state, which we term flocking (dynamic) and which we show as blue region in the state diagram in Fig.\ \ref{fig:2}.

\begin{figure}
	\centering
	\includegraphics[width=12.5cm]{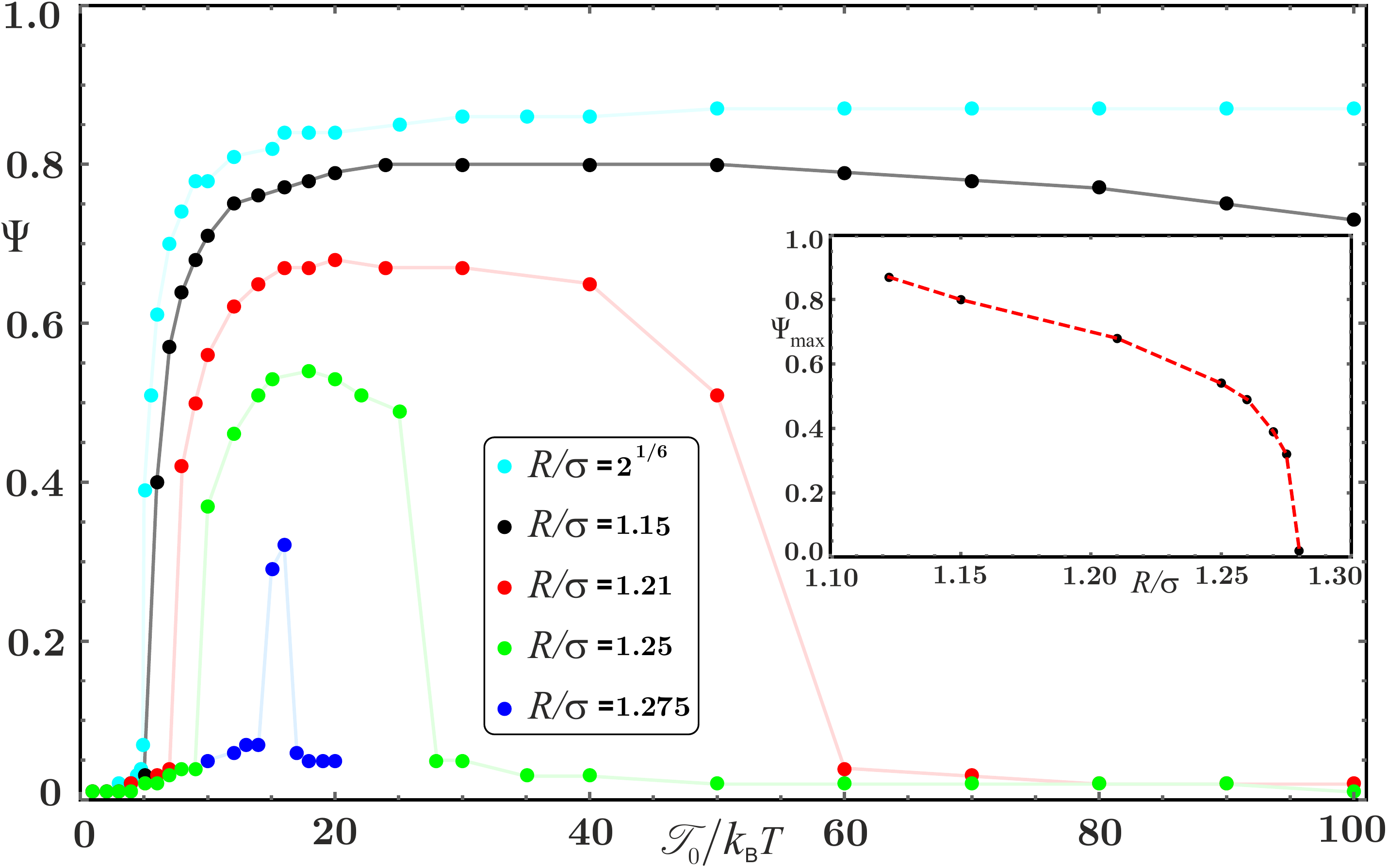}
	\caption{The polar order parameter $\Psi$ as a function of non-reciprocal interaction amplitude $\mathcal{T}_0$ for selected values of interaction range $R$. The inset shows the maximum of the order parameter, $\Psi_\mathrm{max}$, versus $R$. The measured values are shown in (colored) circles. Continuous and dashed lines are guides to the eye.
	}
	\label{fig:4}
\end{figure}

First, to quantify the degree of order in coherently moving states, we use the average normalized orientation $\Psi$ of active particles as the polar order parameter
\begin{equation}
\Psi = \frac{1}{N} \left | \sum_{i=1}^{N} \mathbf{u}_i \right |,
\end{equation} 
where $N$ is the number of particles in the system. If the individual particles point in random directions, the motion is disordered and the order parameter is approximately zero. On the other side, all particles pointing in roughly the same direction would correspond to $\Psi \approx 1$. In Fig. \ref{fig:4} we plot the polar order parameter $\Psi$ as a function of torque amplitude $\mathcal{T}_0$ for several interaction ranges $R$. The highest order $\Psi$ is realized when the range $R$ of non-reciprocal forces coincides with the range of steric forces $2^{1/6}\sigma$. Consistent with the findings of Fig. \ref{fig:2}, the polar order is vanishingly small for $\mathcal{T}_0/k_\mathrm{B}T < 5$ in the MIPS and dispersed state.
Around this torque value, $\Psi$ abruptly increases over a narrow torque interval up to values $\Psi \gtrapprox 0.8$. Interestingly, a structural change in the state of coherent motion from a flocking band to other flocking patterns around $\mathcal{T}_0/k_\mathrm{B}T = 20$ is not accompanied with a change of $\Psi$, which stays approximately the same for a broad range of $\mathcal{T}_0$ values, see Fig. \ref{fig:4}. Increase in the range $R$ of orientational interactions leads to a reduction of order $\Psi$ in general. For sufficiently long interaction range (in our case $R/\sigma \gtrapprox 1.18$) we observe that the order $\Psi$ initially increases with $\mathcal{T}_0$ towards some maximum value, after which $\Psi$ decreases and eventually becomes vanishingly small with a further increase of torque amplitude in the reentrant dispersed state. Intriguingly, for all values of $R$ in this domain, the maximum value of $\Psi$ is roughly attained for torques from the interval $15 < \mathcal{T}_0/k_\mathrm{B}T < 20$. Finally, let us note that the range of torque values for which the system displays coherent motion (finite $\Psi$) decreases with an increase of $R$. For $R/\sigma \ge 1.28$ the order disappears for all $\mathcal{T}_0$. In the inset of Fig.\ \ref{fig:4} we plot the decrease of the maximum of the order parameter, $\Psi_\mathrm{max}$, with the increase of the interaction range $R$. Further work is needed to reveal if the decrease of $\Psi_\mathrm{max}$ to zero in the vicinity of $R/\sigma \approx 1.28$ is continuous or discontinuous.

\begin{figure}
\includegraphics[width=\textwidth]{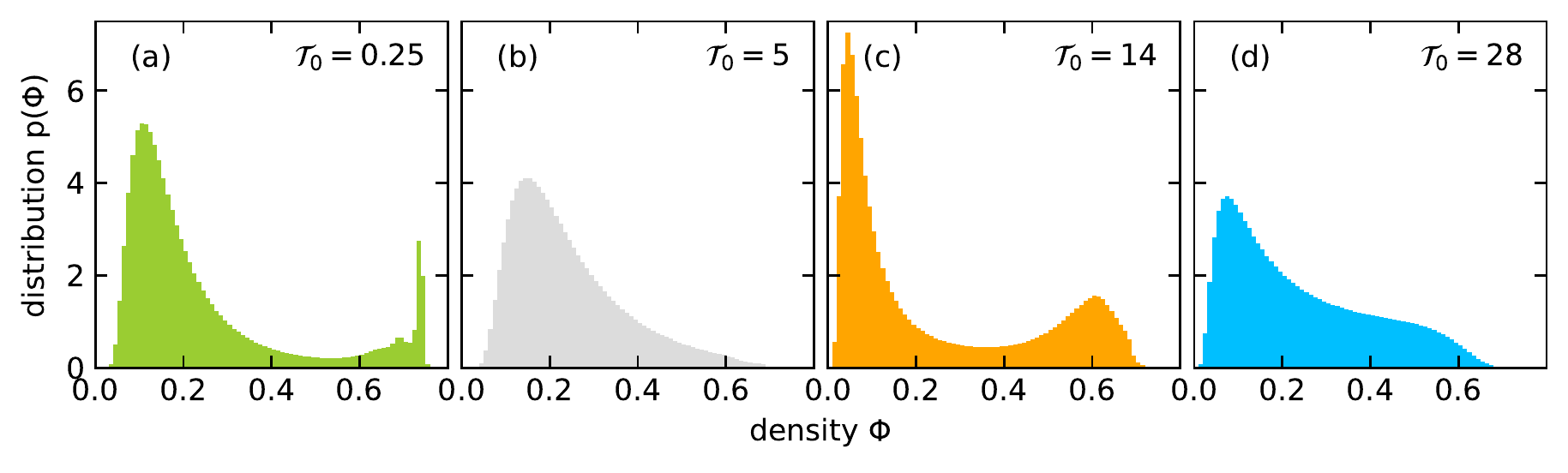}
\caption{Distribution of the local densities, $p(\phi)$, in the four different states for interaction range $R/\sigma = 1.18$ and interaction amplitudes $\mathcal{T}_0$ indicated in each plot. From left to right: MIPS (a), dispersed (b), flocking (band) (c), and flocking (dynamic) (d). The colors of the distributions are used in Fig.\ \ref{fig:6} to indicate the different states.
}
\label{fig:5}
\end{figure}

Second, for snaphots of the different states we perform a Voronoi tesselation of the particle configuration. From the local area $A_\mathrm{loc}$ assigned to each particle, we calculate a local packing fraction or density $\phi = \sigma^2\pi/(4A_\mathrm{loc})$. For the whole time series of such snapshots, we then determine the density distribution $p(\phi)$ for a specific parameter set. Examples for the different states are plotted in Fig.\ \ref{fig:5} for $R/\sigma=1.18$ and different interaction amplitudes $\mathcal{T}_0$. While the distribution for the MIPS state in plot (a) shows a low density peak and a very sharp peak at close-packing density (note, the close-packing peak is not at ca. 0.90 as expected for hard disks but rather at ca. 0.75 since we use the WCA potential of Eq.\ (\ref{eq:WCA}), with a potential value of $\epsilon = 100 k_\mathrm{B}T$ at a particle distance of $\sigma$; therefore, the effective particle diameter is larger than $\sigma$), 
the dispersed state is characterized by a broad maximum of $p(\phi)$ at around $\phi = 0.15$ (b). The density distribution of the flocking (band) state again has two peaks as shown in plot (c). However, compared to the MIPS state the maximum at low densities is sharper and the second maximum, which belongs to the moving band, appears at a density smaller than the close-packing value and is broader. Finally, in the distribution for the flocking (dynamic) state in plot (d) the low-density peak is much less pronounced and the stripes are indicated by a broad shoulder or a very weak maximum (not shown).

\begin{figure}

\centering
\includegraphics[width=10cm]{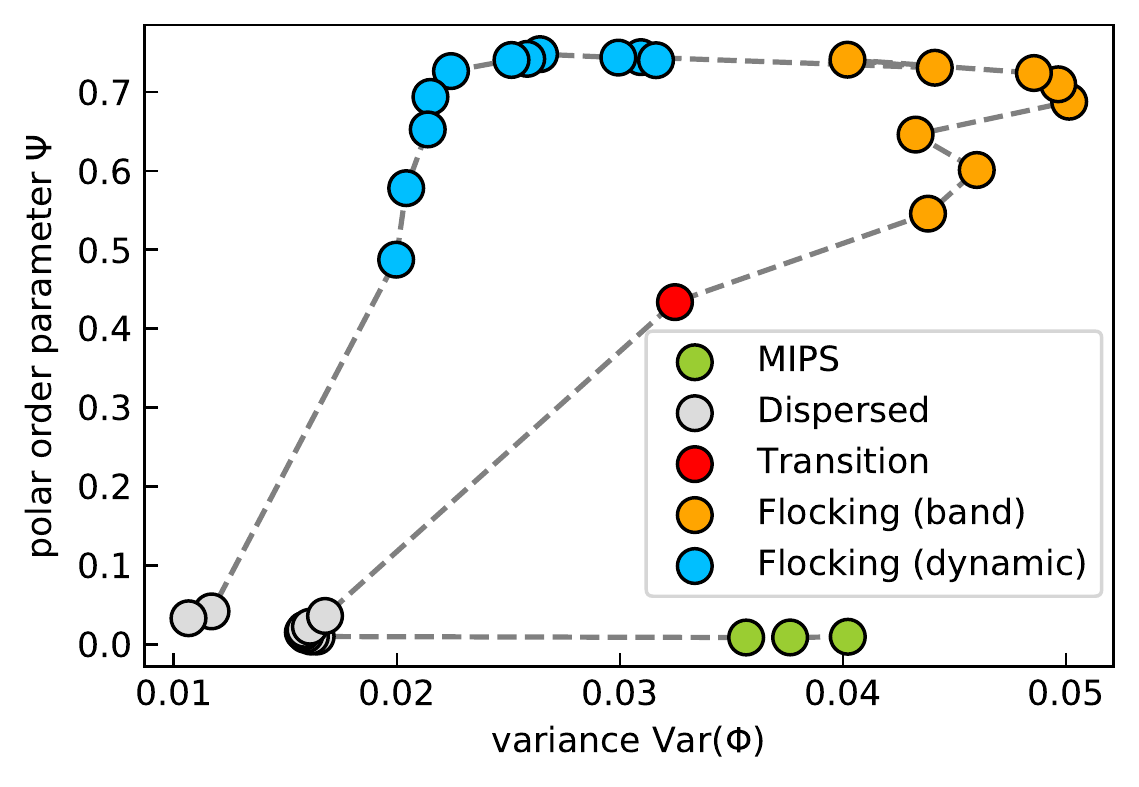}
\caption{Polar order parameter $\Psi$ versus variance $\mathrm{Var}(\phi)$ of the density distributions for all values of $\mathcal{T}_0$ indicated in Fig.\ \ref{fig:2} with interaction range $R/\sigma = 1.18$. The dashed line connects points with increasing value $\mathcal{T}_0$ starting from the MIPS state.
}
\label{fig:6}
\end{figure}

We can now justify our classification of states quantitatively. In Fig.\ \ref{fig:6} we plot the polar order parameter $\Psi$ versus the variance $\mathrm{Var}(\phi)$ of the density distribution for constant interaction range $R/\sigma = 1.18$ and all the maximal torque values $\mathcal{T}_0$ from Fig.\ \ref{fig:2}. The points clearly group into the four states, already introduced. The flocking states have a large polar order parameter, while the compact bands and the dynamic stripes distinguish themselves by the variance of the density distribution, which obviously is larger for the bands. There is only one system at the transition from dispersed to flocking (band) ($\mathcal{T}_0 = 7 k_\mathrm{B}T$), which does not clearly fall into the latter state. The dispersed state and MIPS have nearly zero polar order, as already discussed, but MIPS has a larger variance due to the sharp peak at close-packing density.

Finally, we also checked the occurence of the flocking states for two additional density values  $\phi = 0.15$ and $0.35$ and we also wanted to know if the interaction range, where flocking occurs, expands. Since a complete scan of the parameter space is very time consuming, we chose the specific torque amplitude $\mathcal{T}_0 = 15 k_\mathrm{B}T$, where flocking occurs at the maximum interaction range of $R/\sigma = 1.275$. For the higher density flocking sets in around $R/\sigma = 1.3$ and for the lower density below $R/\sigma = 1.25$. So, there is not a very pronounced variation in the maximum interaction range.

\section*{Discussion}

To summarize, we studied a system of active particles interacting with each other via conventional steric interactions; besides, we introduced the additional non-reciprocal avoidant orientational interactions. The latter were motivated by an example of flagellum--body interactions which are present in an encounter of two algae (cf. Fig.\ \ref{fig:1}). We showed that motility-induced active particle clustering is disturbed already for moderately low non-reciprocal interaction amplitudes $\mathcal{T}_0$ and that the active bath finds itself in most instances in a dispersed state, characterized by numerous and unstable smaller clusters. However, the dispersed state is not the only possible outcome. Remarkably, if the non-reciprocal interaction range $R$ is close to the steric interaction range, we demonstrated that the active system displays coherent flocking motion for a finite range of amplitudes $\mathcal{T}_0$ (cf. Fig.\ \ref{fig:4}). The polar order is found to decrease with increasing $R$, and is largely independent on the underlying flock structure. Together with the variance of the local-density distribution, we can group the flocking motion into band and dynamic stripe states.

The occurrence of a band of flocking particles can at least qualitatively be understood as follows. When they all move in the same direction, the torques acting on each particle from its neighbors cancel. Now, whenever a particle orientation deviates from the direction of the moving band, the particle moves towards its direct neighbor and experiences the avoidant torque back to the moving direction. The same happens when the close-by neighbor on the other side is approached. Thus the parallel orientations of all the particles are stabilized by the avoidant orientational interactions with the neighbors; however, only when the torque amplitude exceeds $k_\mathrm{B}T$ so that thermal orientational motion is irrelevant.
If the interaction range $R$ is too large, a particle experiences the avoidant torques from all the
neighbors simultaneously. The torques cancel each other and the particle's orientation is not turned back to the common direction. Thus, the band disintegrates. Interestingly, it is also not stable for large torques and the reason for this instability under increased ``orientational tension'' requires further explanation. Thus, fully uncovering the microscopic mechanisms of the observed behavior will be one direction of our future efforts.

Collective behavior of \textit{Chlamydomonas} algae has recently been the focus of experimental studies\cite{fragkopoulos:21}. Importantly, the flocking motion observed in our simulation work has not been reported in the experiments so far. We hope that our work may stimulate future investigations aimed at establishing a connection between our parameters $\mathcal{T}_0$ and $R$ and interaction parameters of real systems. A possible experimental setup could be perhaps realized either through biological engineering (manipulating the length and mechanical properties of algal flagella) or by manufacturing artificial microswimmers following prescribed non-reciprocal interactions.   

The non-reciprocal avoidant torque of Eq.\ (\ref{eq:nonrec}) has a very generic form, which one can modify. In particular, there is a symmetry in the torque values for a particle oriented towards or away from its neighbor. One can break this symmetry and make the avoidant torque larger when the particle is oriented towards its neighbor. Such a modification might, for example, be interesting to describe the social interaction of a prey turning away from a predator.

\section*{Methods}

We consider a system consisting of $N$ overdamped active particles propelling with a fixed speed $v$ and variable orientations in two spatial dimensions. The time evolution of their positions $\mathbf{r}_i = (x_i, y_i)$ and orientations $\mathbf{u}_i = (\cos\varphi_i, \sin\varphi_i)$ is described by associated overdamped Langevin equations
\begin{eqnarray}
\dot{\mathbf{r}}_i &=& v \mathbf{u}_i + \mu \sum_{j \neq i} \mathbf{F}_{ij}  + \sqrt{2D}\bm{\xi}_i, \label{eq:active1} \\
\dot{\varphi}_i &=& \mu_\mathrm{R} \sum_{j \neq i} \pmb{\mathcal{T}}_{ij} + \sqrt{2D_{\mathrm{R}}} \eta_i. \label{eq:active2}
\end{eqnarray}
The particles possess translational and rotational mobilities $\mu$ and $\mu_\mathrm{R}$ and their translational and rotational diffusive motion is characterized by respective diffusion constants $D = \mu k_\mathrm{B}T$ and $D_\mathrm{R} = \mu_\mathrm{R} k_\mathrm{B}T$, with $T$ being the ambient temperature. Here $\bm{\xi}_i$ and $\eta_i$ denote two zero-mean and unit-variance Gaussian white noises, $\langle \xi^\alpha_i(t) \xi^\beta_j(t') \rangle = \delta_{\alpha \beta}\delta_{ij}\delta(t-t')$ and $\langle \eta_i(t) \eta_j(t') \rangle = \delta_{ij} \delta(t-t')$, where Greek letters denote Cartesian components. The particles interact with each other in two ways.

Firstly, the particles exhibit steric forces $\mathbf{F}_{ij} = - \nabla_{\mathbf{r}_i} V_{\varepsilon,\sigma}(\mathbf{r}_i - \mathbf{r}_j)$, originating from the Weeks--Chandler--Andersen potential $V_{\epsilon,\sigma}(\mathbf{r})$, which has strength $\varepsilon$ at a characteristic distance $\sigma$: 
\begin{equation}
V_{\varepsilon,\sigma}(\mathbf{r}) = \left\{
\begin{array}{ll}
4 \varepsilon \left [ \left (\frac{\sigma}{|\mathbf{r}|} \right )^{12} - \left (\frac{\sigma}{|\mathbf{r}|} \right )^{6}\right ] + \varepsilon, & \quad |\mathbf{r}| \leq 2^{1/6} \sigma, \\
0, & \quad |\mathbf{r}| > 2^{1/6} \sigma.
\end{array}
\right.
\label{eq:WCA}
\end{equation}
Secondly, the particles located within a distance $R$ from each other experience pairwise orientational interactions. These interactions are modeled by subjecting the particle $i$, when it is close to particle $j$, to the torque
\begin{equation}
\pmb{\mathcal{T}}_{ij} = \mathcal{T}_0 \theta(R - |\mathbf{r}_{ij}|) \left ( \mathbf{u}_i \times \frac{\mathbf{r}_{ij}}{|\mathbf{r}_{ij}|} \right ),
\label{eq:nonrec}
\end{equation}
with $\mathcal{T}_0$ being the torque amplitude, $R$ the interaction range, $\mathbf{r}_{ij} = \mathbf{r}_i - \mathbf{r}_j$ and $\theta(r)$ the Heaviside step function. Note that the orientational interaction is non-reciprocal, $\pmb{\mathcal{T}}_{ij} \neq -\pmb{\mathcal{T}}_{ji}$, in the general case. Some representative examples of this interaction are sketched in Fig. \ref{fig:1}.

We measure time in units of active particle reorientation time $D_\mathrm{R}^{-1}$, length in units of $\sigma$, and energy in units of $k_\mathrm{B}T$. Particle motion is simulated within a box of area $A$ and is subjected to periodic boundary conditions. In the absence of orientational interactions, the state of the active bath is described by the area packing fraction of disks $\Phi = N\sigma^2\pi/(4A)$ and the P\'eclet number $\text{Pe} = \sigma v/D$.

We simulate $N=10^4$ particles, and set $\varepsilon/k_\mathrm{B}T=100$ and $\mu/\mu_\mathrm{R} = \sigma^2/3$, derived from the Stokes friction coefficients of spherical particles. The dimensionless parameters $\Phi = 0.24$ and $\text{Pe} = 80$ are chosen such that the bath exhibits motility-induced phase separation in the absence of orientational interactions, $\mathcal{T}_0 = 0$. The amplitude, $\mathcal{T}_0$, and range, $R$ of the orientational interaction are then varied to explore the state diagram for fixed $\Phi$ and $\text{Pe}$. The numerical integration of Eqs. (\ref{eq:active1}) -- (\ref{eq:active2}) is carried out following an Euler scheme with a time step of $10^{-5} \, D_\mathrm{R}^{-1}$. Upon reaching the steady-state, the data are averaged over 5 independent simulation runs, each of duration $1000 \, D_\mathrm{R}^{-1}$.

\section*{Acknowledgements}
We acknowledge helpful discussion with Oliver B\" aumchen, Marco Mazza and J\'er\'emy Vachier in the initial stage of the research project.
M.K. acknowledges financial support from TU Berlin through a visiting lectureship.

\section*{Author contributions statement}

M.K. and H.S. designed research. M.K. and T.W. carried out numerical simulations and analyzed data. M.K. and H.S. wrote the article.

\section*{Competing interests}

The authors declare no competing interests.

\section*{Data availability}

The datasets used and/or analysed during the current study available from the corresponding author on reasonable request.


\begin{thebibliography}{10}
	\urlstyle{rm}
	\expandafter\ifx\csname url\endcsname\relax
	\def\url#1{\texttt{#1}}\fi
	\expandafter\ifx\csname urlprefix\endcsname\relax\def\urlprefix{URL }\fi
	\expandafter\ifx\csname doiprefix\endcsname\relax\def\doiprefix{DOI: }\fi
	\providecommand{\bibinfo}[2]{#2}
	\providecommand{\eprint}[2][]{\url{#2}}
	
	\bibitem{ramaswamy:10}
	\bibinfo{author}{Ramaswamy, S.}
	\newblock \bibinfo{journal}{\bibinfo{title}{The mechanics and statistics of
			active matter}}.
	\newblock {\emph{\JournalTitle{Annu. Rev. Condens. Matter Phys.}}}
	\textbf{\bibinfo{volume}{1}}, \bibinfo{pages}{323--345}
	(\bibinfo{year}{2010}).
	
	\bibitem{vicsek:12}
	\bibinfo{author}{Vicsek, T.} \& \bibinfo{author}{Zafeiris, A.}
	\newblock \bibinfo{journal}{\bibinfo{title}{Collective motion}}.
	\newblock {\emph{\JournalTitle{Phys. Rep.}}} \textbf{\bibinfo{volume}{517}},
	\bibinfo{pages}{71--140} (\bibinfo{year}{2012}).
	
	\bibitem{elgeti:15}
	\bibinfo{author}{Elgeti, J.}, \bibinfo{author}{Winkler, R.~G.} \&
	\bibinfo{author}{Gompper, G.}
	\newblock \bibinfo{journal}{\bibinfo{title}{Physics of microswimmers -- single
			particle motion and collective behavior: {A} review}}.
	\newblock {\emph{\JournalTitle{Rep. Prog. Phys.}}}
	\textbf{\bibinfo{volume}{78}}, \bibinfo{pages}{056601}
	(\bibinfo{year}{2015}).
	
	\bibitem{zoettl:16}
	\bibinfo{author}{Z\"ottl, A.} \& \bibinfo{author}{Stark, H.}
	\newblock \bibinfo{journal}{\bibinfo{title}{Emergent behavior in active
			colloids}}.
	\newblock {\emph{\JournalTitle{J. Phys.: Condens. Matter}}}
	\textbf{\bibinfo{volume}{28}}, \bibinfo{pages}{253001}
	(\bibinfo{year}{2016}).
	
	\bibitem{bechinger:16}
	\bibinfo{author}{Bechinger, C.}, \bibinfo{author}{Di~Leonardo, R.},
	\bibinfo{author}{L\"owen, H.}, \bibinfo{author}{Volpe, G.} \&
	\bibinfo{author}{Volpe, G.}
	\newblock \bibinfo{journal}{\bibinfo{title}{Active particles in complex and
			crowded environments}}.
	\newblock {\emph{\JournalTitle{Rev. Mod. Phys.}}}
	\textbf{\bibinfo{volume}{88}}, \bibinfo{pages}{045006}
	(\bibinfo{year}{2016}).
	
	\bibitem{romanczuk:12}
	\bibinfo{author}{Romanczuk, P.}, \bibinfo{author}{B\"ar, M.},
	\bibinfo{author}{Ebeling, W.}, \bibinfo{author}{Lindner, B.} \&
	\bibinfo{author}{Schimansky-Geier, L.}
	\newblock \bibinfo{journal}{\bibinfo{title}{Active {Brownian} particles}}.
	\newblock {\emph{\JournalTitle{Eur. Phys. J. Spec. Top.}}}
	\textbf{\bibinfo{volume}{202}}, \bibinfo{pages}{1--162}
	(\bibinfo{year}{2012}).
	
	\bibitem{tailleur:08}
	\bibinfo{author}{Tailleur, J.} \& \bibinfo{author}{Cates, M.~E.}
	\newblock \bibinfo{journal}{\bibinfo{title}{Statistical mechanics of
			interacting run-and-tumble bacteria}}.
	\newblock {\emph{\JournalTitle{Phys. Rev. Lett.}}}
	\textbf{\bibinfo{volume}{100}}, \bibinfo{pages}{218103}
	(\bibinfo{year}{2008}).
	
	\bibitem{filly:12}
	\bibinfo{author}{Filly, Y.} \& \bibinfo{author}{Marchetti, M.~C.}
	\newblock \bibinfo{journal}{\bibinfo{title}{Athermal phase separation of
			self-propelled particles with no alignment}}.
	\newblock {\emph{\JournalTitle{Phys. Rev. Lett.}}}
	\textbf{\bibinfo{volume}{108}}, \bibinfo{pages}{235702}
	(\bibinfo{year}{2012}).
	
	\bibitem{redner:13}
	\bibinfo{author}{Redner, G.~S.}, \bibinfo{author}{Hagan, M.~F.} \&
	\bibinfo{author}{Baskaran, A.}
	\newblock \bibinfo{journal}{\bibinfo{title}{Structure and dynamics of a
			phase-separating active colloidal fluid}}.
	\newblock {\emph{\JournalTitle{Phys. Rev. Lett.}}}
	\textbf{\bibinfo{volume}{110}}, \bibinfo{pages}{055701}
	(\bibinfo{year}{2013}).
	
	\bibitem{cates:15}
	\bibinfo{author}{Cates, M.~E.} \& \bibinfo{author}{Tailleur, J.}
	\newblock \bibinfo{journal}{\bibinfo{title}{Motility-induced phase
			separation}}.
	\newblock {\emph{\JournalTitle{Annu. Rev. Condens. Matter Phys.}}}
	\textbf{\bibinfo{volume}{6}}, \bibinfo{pages}{219--244}
	(\bibinfo{year}{2015}).
	
	\bibitem{buttinoni:13}
	\bibinfo{author}{Buttinoni, I.} \emph{et~al.}
	\newblock \bibinfo{journal}{\bibinfo{title}{Dynamical clustering and phase
			separation in suspensions of self-propelled colloidal particles}}.
	\newblock {\emph{\JournalTitle{Phys. Rev. Lett.}}}
	\textbf{\bibinfo{volume}{110}}, \bibinfo{pages}{238301}
	(\bibinfo{year}{2013}).
	
	\bibitem{speck:16}
	\bibinfo{author}{Speck, T.}
	\newblock \bibinfo{journal}{\bibinfo{title}{Collective behavior of active
			{Brownian} particles: {From} microscopic clustering to macroscopic phase
			separation}}.
	\newblock {\emph{\JournalTitle{Eur. Phys. J. Special Topics}}}
	\textbf{\bibinfo{volume}{225}}, \bibinfo{pages}{2287--2299}
	(\bibinfo{year}{2016}).
	
	\bibitem{lavergne:19}
	\bibinfo{author}{Lavergne, F.~A.}, \bibinfo{author}{Wendehenne, H.},
	\bibinfo{author}{B\"auerle, T.} \& \bibinfo{author}{Bechinger, C.}
	\newblock \bibinfo{journal}{\bibinfo{title}{Group formation and cohesion of
			active particles with visual perception-dependent motility}}.
	\newblock {\emph{\JournalTitle{Science}}} \textbf{\bibinfo{volume}{364}},
	\bibinfo{pages}{70--74} (\bibinfo{year}{2019}).
	
	\bibitem{baeuerle:20}
	\bibinfo{author}{B\"auerle, T.}, \bibinfo{author}{L\"offler, R.~C.} \&
	\bibinfo{author}{Bechinger, C.}
	\newblock \bibinfo{journal}{\bibinfo{title}{Formation of stable and responsive
			collective states in suspensions of active colloids}}.
	\newblock {\emph{\JournalTitle{Nat. Commun.}}} \textbf{\bibinfo{volume}{11}},
	\bibinfo{pages}{2547} (\bibinfo{year}{2020}).
	
	\bibitem{vicsek:95}
	\bibinfo{author}{Vicsek, T.}, \bibinfo{author}{Czir\'ok, A.},
	\bibinfo{author}{Ben-Jacob, E.}, \bibinfo{author}{Cohen, I.} \&
	\bibinfo{author}{Shochet, O.}
	\newblock \bibinfo{journal}{\bibinfo{title}{Novel type of phase transition in a
			system of self-driven particles}}.
	\newblock {\emph{\JournalTitle{Phys. Rev. Lett.}}}
	\textbf{\bibinfo{volume}{75}}, \bibinfo{pages}{1226--1229}
	(\bibinfo{year}{1995}).
	
	\bibitem{toner:95}
	\bibinfo{author}{Toner, J.} \& \bibinfo{author}{Tu, Y.}
	\newblock \bibinfo{journal}{\bibinfo{title}{Long-range order in a
			two-dimensional dynamical {XY} model: {How} birds fly together}}.
	\newblock {\emph{\JournalTitle{Phys. Rev. Lett.}}}
	\textbf{\bibinfo{volume}{75}}, \bibinfo{pages}{4326--4329}
	(\bibinfo{year}{1995}).
	
	\bibitem{czirok:97}
	\bibinfo{author}{Czir\'ok, A.}, \bibinfo{author}{Stanley, H.~E.} \&
	\bibinfo{author}{Vicsek, T.}
	\newblock \bibinfo{journal}{\bibinfo{title}{Spontaneously ordered motion of
			self-propelled particles}}.
	\newblock {\emph{\JournalTitle{J. Phys. A: Math. Gen.}}}
	\textbf{\bibinfo{volume}{30}}, \bibinfo{pages}{1375--1385}
	(\bibinfo{year}{1997}).
	
	\bibitem{toner:98}
	\bibinfo{author}{Toner, J.} \& \bibinfo{author}{Tu, Y.}
	\newblock \bibinfo{journal}{\bibinfo{title}{Flocks, herds, and schools: {A}
			quantitative theory of flocking}}.
	\newblock {\emph{\JournalTitle{Phys. Rev. E}}} \textbf{\bibinfo{volume}{58}},
	\bibinfo{pages}{4828--4858} (\bibinfo{year}{1998}).
	
	\bibitem{gregoire:04}
	\bibinfo{author}{Gr\'egoire, G.} \& \bibinfo{author}{Chat\'e, H.}
	\newblock \bibinfo{journal}{\bibinfo{title}{Onset of collective and cohesive
			motion}}.
	\newblock {\emph{\JournalTitle{Phys. Rev. Lett.}}}
	\textbf{\bibinfo{volume}{92}}, \bibinfo{pages}{025702}
	(\bibinfo{year}{2004}).
	
	\bibitem{martin-gomez:18}
	\bibinfo{author}{Mart\'in-G\'omez, A.}, \bibinfo{author}{Levis, D.},
	\bibinfo{author}{D\'iaz-Guilera, A.} \& \bibinfo{author}{Pagonabarraga, I.}
	\newblock \bibinfo{journal}{\bibinfo{title}{Collective motion of active
			{Brownian} particles with polar alignment}}.
	\newblock {\emph{\JournalTitle{Soft Matter}}} \textbf{\bibinfo{volume}{14}},
	\bibinfo{pages}{2610--2618} (\bibinfo{year}{2018}).
	
	\bibitem{Zhang:21}
	\bibinfo{author}{Zhang, J.}, \bibinfo{author}{Alert, R.}, \bibinfo{author}{Yan,
		J.}, \bibinfo{author}{Wingreen, N.~S.} \& \bibinfo{author}{S., G.}
	\newblock \bibinfo{journal}{\bibinfo{title}{Active phase separation by turning
			towards regions of higher density}}.
	\newblock {\emph{\JournalTitle{Nature Phys.}}} \textbf{\bibinfo{volume}{17}},
	\bibinfo{pages}{961} (\bibinfo{year}{2021}).
	
	\bibitem{Linden:19}
	\bibinfo{author}{Van Der~Linden, M.~N.}, \bibinfo{author}{Alexander, L.~C.},
	\bibinfo{author}{Aarts, D.~G.} \& \bibinfo{author}{Dauchot, O.}
	\newblock \bibinfo{journal}{\bibinfo{title}{Interrupted motility induced phase
			separation in aligning active colloids}}.
	\newblock {\emph{\JournalTitle{Phys. Rev. Lett.}}}
	\textbf{\bibinfo{volume}{123}}, \bibinfo{pages}{098001}
	(\bibinfo{year}{2019}).
	
	\bibitem{Pu:17}
	\bibinfo{author}{Pu, M.}, \bibinfo{author}{Jiang, H.} \& \bibinfo{author}{Hou,
		Z.}
	\newblock \bibinfo{journal}{\bibinfo{title}{Reentrant phase separation behavior
			of active particles with anisotropic janus interaction}}.
	\newblock {\emph{\JournalTitle{Soft Matter}}} \textbf{\bibinfo{volume}{13}},
	\bibinfo{pages}{4112} (\bibinfo{year}{2017}).
	
	\bibitem{Liao:20}
	\bibinfo{author}{Liao, G.-J.}, \bibinfo{author}{Hall, C.~K.} \&
	\bibinfo{author}{Klapp, S.~H.}
	\newblock \bibinfo{journal}{\bibinfo{title}{Dynamical self-assembly of dipolar
			active brownian particles in two dimensions}}.
	\newblock {\emph{\JournalTitle{Soft Matter}}} \textbf{\bibinfo{volume}{16}},
	\bibinfo{pages}{2208} (\bibinfo{year}{2020}).
	
	\bibitem{Barre:15}
	\bibinfo{author}{Barr{\'e}, J.}, \bibinfo{author}{Ch{\'e}trite, R.},
	\bibinfo{author}{Muratori, M.} \& \bibinfo{author}{Peruani, F.}
	\newblock \bibinfo{journal}{\bibinfo{title}{Motility-induced phase separation
			of active particles in the presence of velocity alignment}}.
	\newblock {\emph{\JournalTitle{J. Stat. Phys.}}}
	\textbf{\bibinfo{volume}{158}}, \bibinfo{pages}{589} (\bibinfo{year}{2015}).
	
	\bibitem{Sansa:18}
	\bibinfo{author}{Sese-Sansa, E.}, \bibinfo{author}{Pagonabarraga, I.} \&
	\bibinfo{author}{Levis, D.}
	\newblock \bibinfo{journal}{\bibinfo{title}{Velocity alignment promotes
			motility-induced phase separation}}.
	\newblock {\emph{\JournalTitle{EPL}}} \textbf{\bibinfo{volume}{124}},
	\bibinfo{pages}{30004} (\bibinfo{year}{2018}).
	
	\bibitem{Bhatta:19}
	\bibinfo{author}{Bhattacherjee, B.} \& \bibinfo{author}{Chaudhuri, D.}
	\newblock \bibinfo{journal}{\bibinfo{title}{Re-entrant phase separation in
			nematically aligning active polar particles}}.
	\newblock {\emph{\JournalTitle{Soft Matter}}} \textbf{\bibinfo{volume}{15}},
	\bibinfo{pages}{8483} (\bibinfo{year}{2019}).
	
	\bibitem{Mallory:19}
	\bibinfo{author}{Mallory, S.~A.} \& \bibinfo{author}{Cacciuto, A.}
	\newblock \bibinfo{journal}{\bibinfo{title}{Activity-enhanced self-assembly of
			a colloidal kagome lattice}}.
	\newblock {\emph{\JournalTitle{JACS}}} \textbf{\bibinfo{volume}{141}},
	\bibinfo{pages}{2500} (\bibinfo{year}{2019}).
	
	\bibitem{ivlev:15}
	\bibinfo{author}{Ivlev, A.~V.} \emph{et~al.}
	\newblock \bibinfo{journal}{\bibinfo{title}{Statistical mechanics where
			{Newton's} thrid law is broken}}.
	\newblock {\emph{\JournalTitle{Phys. Rev. X}}} \textbf{\bibinfo{volume}{5}},
	\bibinfo{pages}{011035} (\bibinfo{year}{2015}).
	
	\bibitem{uchida:10}
	\bibinfo{author}{Uchida, N.} \& \bibinfo{author}{Golestanian, R.}
	\newblock \bibinfo{journal}{\bibinfo{title}{Synchronization and collective
			dynamics in a carpet of microfluidic rotors}}.
	\newblock {\emph{\JournalTitle{Phys. Rev. Lett.}}}
	\textbf{\bibinfo{volume}{104}}, \bibinfo{pages}{178103}
	(\bibinfo{year}{2010}).
	
	\bibitem{meredith:20}
	\bibinfo{author}{Meredith, C.~H.} \emph{et~al.}
	\newblock \bibinfo{journal}{\bibinfo{title}{Predator-prey interactions between
			droplets driven by non-reciprocal oil exchange}}.
	\newblock {\emph{\JournalTitle{Nature Chemistry}}}
	\textbf{\bibinfo{volume}{12}}, \bibinfo{pages}{1136--1142}
	(\bibinfo{year}{2020}).
	
	\bibitem{sompolinsky:86}
	\bibinfo{author}{Sompolinsky, H.} \& \bibinfo{author}{Kanter, I.}
	\newblock \bibinfo{journal}{\bibinfo{title}{Temporal association in asymmetric
			neural networks}}.
	\newblock {\emph{\JournalTitle{Phys. Rev. Lett.}}}
	\textbf{\bibinfo{volume}{57}}, \bibinfo{pages}{2861--2864}
	(\bibinfo{year}{1986}).
	
	\bibitem{montbrio:18}
	\bibinfo{author}{Montbri\'o, E.} \& \bibinfo{author}{Paz\'o, D.}
	\newblock \bibinfo{journal}{\bibinfo{title}{Kuramoto model for
			excitation-inhibition-based oscillations}}.
	\newblock {\emph{\JournalTitle{Phys. Rev. Lett.}}}
	\textbf{\bibinfo{volume}{120}}, \bibinfo{pages}{244101}
	(\bibinfo{year}{2018}).
	
	\bibitem{fruchart:21}
	\bibinfo{author}{Fruchart, M.}, \bibinfo{author}{Hanai, R.},
	\bibinfo{author}{Littlewood, P.~B.} \& \bibinfo{author}{Vitelli, V.}
	\newblock \bibinfo{journal}{\bibinfo{title}{Non-reciprocal phase transitions}}.
	\newblock {\emph{\JournalTitle{Nature}}} \textbf{\bibinfo{volume}{592}},
	\bibinfo{pages}{363--369} (\bibinfo{year}{2021}).
	
	\bibitem{soto:14}
	\bibinfo{author}{Soto, R.} \& \bibinfo{author}{Golestanian, R.}
	\newblock \bibinfo{journal}{\bibinfo{title}{Self-assembly of catalytically
			active colloidal molecules: {Tailoring} activity through surface chemistry}}.
	\newblock {\emph{\JournalTitle{Phys. Rev. Lett.}}}
	\textbf{\bibinfo{volume}{112}}, \bibinfo{pages}{068301}
	(\bibinfo{year}{2014}).
	
	\bibitem{pohl:14}
	\bibinfo{author}{Pohl, O.} \& \bibinfo{author}{Stark, H.}
	\newblock \bibinfo{journal}{\bibinfo{title}{Dynamic clustering and chemotactic
			collapse of self-phoretic active particles}}.
	\newblock {\emph{\JournalTitle{Phys. Rev. Lett.}}}
	\textbf{\bibinfo{volume}{112}}, \bibinfo{pages}{238303}
	(\bibinfo{year}{2014}).
	
	\bibitem{saha:19}
	\bibinfo{author}{Saha, S.}, \bibinfo{author}{Ramaswamy, S.} \&
	\bibinfo{author}{Golestanian, R.}
	\newblock \bibinfo{journal}{\bibinfo{title}{Pairing, waltzing and scattering of
			chemotactic active colloids}}.
	\newblock {\emph{\JournalTitle{New. J. Phys.}}} \textbf{\bibinfo{volume}{21}},
	\bibinfo{pages}{063006} (\bibinfo{year}{2019}).
	
	\bibitem{stuermer:19}
	\bibinfo{author}{St\"urmer, J.}, \bibinfo{author}{Seyrich, M.} \&
	\bibinfo{author}{Stark, H.}
	\newblock \bibinfo{journal}{\bibinfo{title}{Chemotaxis in a binary mixture of
			active and passive particles}}.
	\newblock {\emph{\JournalTitle{J. Chem. Phys.}}}
	\textbf{\bibinfo{volume}{150}}, \bibinfo{pages}{214901}
	(\bibinfo{year}{2019}).
	
	\bibitem{harris:09}
	\bibinfo{author}{Harris, E.~H.}
	\newblock \emph{\bibinfo{title}{The {Chlamydomonas} sourcebook}}
	(\bibinfo{publisher}{Academic {Press}}, \bibinfo{address}{Oxford},
	\bibinfo{year}{2009}).
	
	\bibitem{polin:09}
	\bibinfo{author}{Polin, M.}, \bibinfo{author}{Tuval, I.},
	\bibinfo{author}{Drescher, K.}, \bibinfo{author}{Gollub, J.~P.} \&
	\bibinfo{author}{Goldstein, R.~E.}
	\newblock \bibinfo{journal}{\bibinfo{title}{Chlamydomonas swims with two
			"{Gears}" in a {Eukaryotic} version of run-and-tumble locomotion}}.
	\newblock {\emph{\JournalTitle{Science}}} \textbf{\bibinfo{volume}{325}},
	\bibinfo{pages}{487--490} (\bibinfo{year}{2009}).
	
	\bibitem{drescher:10}
	\bibinfo{author}{Drescher, K.}, \bibinfo{author}{Goldstein, R.~E.},
	\bibinfo{author}{Michel, N.}, \bibinfo{author}{Polin, M.} \&
	\bibinfo{author}{Tuval, I.}
	\newblock \bibinfo{journal}{\bibinfo{title}{Direct measurement of the flow
			field around swimming microorganisms}}.
	\newblock {\emph{\JournalTitle{Phys. Rev. Lett.}}}
	\textbf{\bibinfo{volume}{105}}, \bibinfo{pages}{168101}
	(\bibinfo{year}{2010}).
	
	\bibitem{kantsler:13}
	\bibinfo{author}{Kantsler, V.}, \bibinfo{author}{Dunkel, J.},
	\bibinfo{author}{Polin, M.} \& \bibinfo{author}{Goldstein, R.~E.}
	\newblock \bibinfo{journal}{\bibinfo{title}{Ciliary contact interactions
			dominate surface scattering of swimming eukaryotes}}.
	\newblock {\emph{\JournalTitle{Proc. Natl. Acad. Sci.}}}
	\textbf{\bibinfo{volume}{110}}, \bibinfo{pages}{1187--1192}
	(\bibinfo{year}{2013}).
	
	\bibitem{contino:15}
	\bibinfo{author}{Contino, M.}, \bibinfo{author}{Lushi, E.},
	\bibinfo{author}{Tuval, I.}, \bibinfo{author}{Kantsler, V.} \&
	\bibinfo{author}{Polin, M.}
	\newblock \bibinfo{journal}{\bibinfo{title}{Microalgae scatter off solid
			surfaces by hydrodynamic and contact forces}}.
	\newblock {\emph{\JournalTitle{Phys. Rev. Lett.}}}
	\textbf{\bibinfo{volume}{115}}, \bibinfo{pages}{258102}
	(\bibinfo{year}{2015}).
	
	\bibitem{ostapenko:18}
	\bibinfo{author}{Ostapenko, T.} \emph{et~al.}
	\newblock \bibinfo{journal}{\bibinfo{title}{Curvature-guided motility of
			microalgae in geometric confinement}}.
	\newblock {\emph{\JournalTitle{Phys. Rev. Lett.}}}
	\textbf{\bibinfo{volume}{120}}, \bibinfo{pages}{068002}
	(\bibinfo{year}{2018}).
	
	\bibitem{fragkopoulos:21}
	\bibinfo{author}{Fragkopoulos, A.~A.} \emph{et~al.}
	\newblock \bibinfo{journal}{\bibinfo{title}{Self-generated oxygen gradients
			control collective aggregation of photosyntetic microbes}}.
	\newblock {\emph{\JournalTitle{J. R. Soc. Interface}}}
	\textbf{\bibinfo{volume}{18}}, \bibinfo{pages}{20210553}
	(\bibinfo{year}{2021}).
	
\end{thebibliography}
\end{document}